# Vibrational temperature of the adlayer in 'hot atom' reaction mechanism

By

M. Tomellini

*Dipartimento di Scienze e Tecnologie Chimiche Università di Roma Tor Vergata, Via della Ricerca Scientifica 00133 Roma Italy.*

**Abstract**

Hot-atoms reactions mechanisms bring about reaction rates which are several orders of magnitude higher than those expected in the case of ad-atoms which have thermalized with the surface. This paper addresses the issue of a possible thermodynamic characterization of the adlayer under reactive conditions and at the steady state. In turn, this implies to tackle the question of determining the temperature of the ad-atoms. This is done by means of a nonequilibrium statistical thermodynamic approach, by exploiting a suitable definition of the entropy. The interplay between reaction rate, vibrational temperature of the ad-atoms and adsorbed quantities is highlighted. It is shown that the vibrational temperature depends on reaction rate logarithmically and exhibits a non-linear scaling on physical quantities linked to the energetics of the reaction, namely the adsorption energy and the binding energy of the molecule. The present modeling is also discussed in connection with response equations of nonequilibrium thermodynamics.

**Introduction**

During the last decades the study of atom recombination at catalytic surfaces has been an issue of great moment from both experimental and theoretical perspectives [1-11 and ref. therein]. Experimental findings on reaction cross sections which are intermediate between those proper of the Langmuir Hinshelwood (LH) and Eley Rideal (ER) mechanisms, put in evidence the possibility of a different mechanism governed by the so-called 'hot' ad-atoms [12-15]. The adjective 'hot' is referred to an adspecies which, although trapped in the adsorption potential well, has not thermalized with the catalyst surface. These adspecies are expected to be highly reactive on the surface and, for this reason, their role is thought to be fundamental for the kinetics of the process.

For reactions proceeding under steady state conditions determining the rate of the reaction implies the knowledge of how adatoms are distributed in energy, that is their occupation number in the vibrational level of the adsorption potential well. In the case of 'hot-atom' reaction mechanism, these occupation numbers are expected to be grater than those of a Boltzmann distribution at the temperature of the surface; in this respect the energy distribution function (d.f.) of the adatoms is hyperthermal.

Modeling the energy d.f. of adatoms, in chemisorption and catalysis, has been the subject of several works in the literature [16-20]. In these approaches two processes have to be taken into account in computing the d.f., namely the accommodation of the adatom at the adsorption site and the formation of the diatom. The former is linked to the energy dissipation of the atom, trapped in the adsorption potential well during adsorption, to the solid. The latter implies binary collisions among reacting species. An approach suitable for computing the vibrational d.f. rests on kinetic equations where energy dissipation and atom recombination are described as first and second order reactions, respectively. Also, continuum approaches based on the use of the Fokker-Planck equation have been employed in the literature [16, 18]. In particular, in these models gas atoms enter the adsorption potential well in the upper bound level of the vibrational ladder, and come down the ladder owing to the energy transfer to the solid surface and/or to the adlayer. It is the interplay between adsorption and recombination rates and energy disposal to the solid, which affects the vibrational d.f. of the adatoms. At the steady state these modelings lead to analytical solutions for the d.f. which, in turn, have been employed for interpreting experimental data in order to gain information on reaction rate and on the process of energy disposal as well [21].

So far these approaches, being 'kinetic' in nature, were aimed at characterizing the system (substrate-adlayer) through kinetic quantities, such as the reaction rate and the rate coefficients for energy disposal. The purpose of the present paper is to attempt a thermodynamic characterization of the adlayer, under nonequilibrium reaction conditions, at steady state. This is achieved through the computation of the entropy of the adsorbate together with a suitable definition of the temperature of the adlayer. To this end the knowledge of the d.f., for instance computed using the kinetic approach above quoted, is needed. These quantities are further employed to estimate parameters typical of the nonequilibrium thermodynamic equations, in order to bridge the gap between the previously employed kinetic approach and that based on response equation of nonequilibrium thermodynamics.

**Results and discussion**

*1-The model*

In this contribution we consider the surface reaction

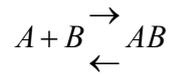

$$A + B \underset{\leftarrow}{\overset{\rightarrow}{}} AB$$

where $A$, $B$ are the adatoms and $AB$ is the gas molecule. The reaction rate is given by

$$\frac{dn_{AB}}{dt} = r = R_d - R_r, \tag{1}$$

that is the difference between direct and reverse reactions. In the following we deal with a mobile adlayer model according to which adatoms are trapped in a 1D potential well (along the normal to the surface) and move freely on the surface plane. By considering the reaction to take place between A and B species in the vibrational ladder of the adsorption potential well, these rates are

$$R_d = \sum_{i,j,k} w_{i_\alpha j_\lambda, k} n_{i_\alpha} n_{j_\lambda}, \tag{2a}$$

$$R_r = \sum_{i,j,k} w_{k, i_\alpha j_\lambda} n_k \tag{2b}$$

where $n$ is the number of atoms (molecules), $w_{i_\alpha j_\lambda, k}$ denotes the state to state transition probability for the open reaction channel $A_{i_\alpha} + B_{j_\lambda} \underset{\leftarrow}{\rightarrow} AB_k$ between an A adatom in vibrational level $i$ and translational state $\alpha$, and a B adatom in its vibrational level $j$, and translational state $\lambda$, to produce a molecule in state $k$. In eqn.2 the sum is performed over the state vectors $\underline{i} \equiv (i_1, i_2, ..., i_m, ...)$. The employed notation emphasizes the vibrational state of the adatoms (i, j). For each reaction channel energy conservation holds, namely $E_k = E_{i_\alpha} + E_{j_\lambda}$. By introducing the probability to find a species in state $i_\alpha$, $p_{i_\alpha} = \dfrac{n_{i_\alpha}}{n}$ where $n = \sum_{\underline{i}} n_{i_\alpha}$ is the total number of species, the reaction rate eqn.1 becomes

$$r = k_d n_A n_B - k_r n_{AB} \qquad (3)$$

where

$$k_d = \sum_{\underline{i}\underline{j},k} w_{i_\alpha j_\lambda, k} p_{i_\alpha} p_{j_\lambda} \qquad (4a)$$

$$k_r = \sum_{k,\underline{i}\underline{j}} w_{k, i_\alpha j_\lambda} p_k \qquad (4b)$$

are the rate coefficients [22]. In eqns.4 the sums are performed over the states of the open reaction channels. Specifically, in these expressions the sums run over the initial and final states and $p_{i_\alpha} \equiv p(E_{i_\alpha})$ is the energy distribution function of the vibrational ladder. The ratio between these two constants reads $\dfrac{k_d}{k_r} = \dfrac{\sum_{\underline{i}\underline{j},k} w_{i_\alpha j_\lambda, k} p_{i_\alpha} p_{j_\lambda}}{\sum_{k,\underline{i}\underline{j}} w_{k, i_\alpha j_\lambda} p_k}$ where for each reaction channel the transition probabilities of direct and reverse processes are equal (microscopic reversibility) : $w_{i_\alpha j_\lambda, k} \equiv w_{k, i_\alpha j_\lambda}$.

Let us first consider the case of a nonequilibrium reaction where the reagents (adspecies) and the product (AB molecules) are in thermal equilibrium at the substrate temperature, $T_s$. Moreover, as far as the adspecies are concerned we deal with a mobile adlayer model where the probability function is $p_{i_\alpha} = \dfrac{e^{-\beta_s E_{i_\alpha}}}{q}$ with $q = \sum_{\underline{i}} e^{-\beta_s E_{i_\alpha}}$ single particle partition function and $\beta_s = 1/k_B T_s$, $k_B$ being the Boltzmann constant. By inserting this expression in the relation above, and exploiting the equality of the transition probabilities, one gets

$$\frac{k_d}{k_r} = \frac{q_{AB}}{q_A q_B} = e^{-\beta_s(\mu_{AB}^0 - \mu_A^0 - \mu_B^0)}, \tag{5}$$

where $\mu_i^0 = -k_B T_s \ln q_i$ is the standard chemical potential and $\mu_i = \mu_i^0 + k_B T_s \ln n_i$ the chemical potential. This expression of the chemical potential is attained by considering independent and indistinguishable particles. Eqn.5 is the detailed balancing i.e. the ratio between the rate constants is equal to the equilibrium constant of the reaction. It is worth noting, in passing, that writing the equilibrium constant of the reaction as $\frac{k_d}{k_r} = \frac{P_{AB}}{\theta_A \theta_B}$ (with $P$ being the gas pressure and $\theta_i$ the surface coverage at equilibrium) would require the following definition of the standard chemical potentials: $\mu'^0_{A(B)} = \mu^0_{A(B)} + k_B T_s \ln M$ and $\mu'^0_{AB} = \mu^0_{AB} + k_B T_s \ln \beta_s V$ where $M$ is the number of adsorption sites and $V$ the total volume of the reactor. Using the chemical potential the reaction rate can be rewritten as

$$r = \left(\sum_{\underline{i},\underline{j},k} w_{i_\alpha j_\lambda, k} e^{-\beta_s(E_{i_\alpha} + E_{j_\lambda})}\right) e^{\beta_s(\mu_A + \mu_B)} - \left(\sum_{\underline{i},\underline{j},k} w_{k, i_\alpha j_\lambda} e^{-\beta_s(E_k)}\right) e^{\beta_s \mu_{AB}}$$
$$= \left(\sum_{\underline{i},\underline{j},k} w_{i_\alpha j_\lambda, k} e^{-\beta_s(E_{i_\alpha} + E_{j_\lambda})}\right) e^{\beta_s(\mu_A + \mu_B)} (1 - e^{\beta_s A_r}) = R_d (1 - e^{\beta_s A_r}) \tag{6}$$

where $A_r = \mu_{AB} - \mu_A - \mu_B$ is the affinity of the reaction. In the case of small departure from equilibrium eqn.6 leads to the "kinetic law of mass action" [23, 24] which implies $r \propto -A_r$. It goes without saying that when the direct reaction prevails (for instance when the product are continually removed), $A_r \to -\infty$ in eqn.6 and $r = R_d > 0$. Furthermore, if equilibrium is established between adsorbed species and the same component in the gas phase, then the chemical potentials entering eqn.6 are equal to those of the atoms in the gas phase.

Let us now pass to discuss the nonequilibrium case where adspecies have not thermalised with the surface in the course of the recombination reaction at steady state. Under these circumstances differences arise with the situation previously studied, ascribed to both the rate constant and the determination of the adsorbed quantities (surface coverage). The rate constant now depends on the probability function $p_{i_\alpha}$, which is not given by the Boltzmann distribution anymore; in turn, as discussed in more detail below, it is expected to depend upon reaction rate. As regards the surface coverage, while in the equilibrium case it can be linked to thermodynamic quantities (such as the chemical potential), under nonequilibrium conditions it is determined by the reaction kinetics. For a non-Boltzmannian energy distribution function of the vibrational ladder it is not straightforward to define the adlayer temperature, although the mean energy and the entropy of the adlayer can be both defined. In fact, the entropy of the adlayer can be estimated by exploiting the definition proposed in [25, 26] which also holds in the nonequilibrium case

$$s = -k_B[\sum_i p_{i_\alpha} \ln p_{i_\alpha} + \ln n], \tag{7}$$

$s$ being the partial molar entropy. In order to write the reaction rate through an expression similar to eqn.6 (as usually employed in nonequilibrium thermodynamics) one defines the quantity $\mu' = \overline{E} - T_s s = \sum_i E_{i_\alpha} p_{i_\alpha} + k_B T_s \sum_i p_{i_\alpha} \ln p_{i_\alpha} + k_B T_s \ln n$, which resembles the thermodynamic expression of the chemical potential. (Please notice that in this definition the term $-a\frac{\partial(\overline{E}-T_s s)}{\partial a}$, where $a = A/n$ is the area per particle, has been omitted. This is in view of the presentation that follows where only vibrational states are taken into account, and the vibrational frequency is independent of $a$). In the equation above $\overline{E} = \sum_i p_{i_\alpha} E_{i_\alpha}$ is the mean energy of the adatom. However, one has to bear in mind that in the present context $\mu'$ is just a mathematical definition suitable for expressing the rate in the form of eqn.6. Nevertheless, for it is possible to define a vibrational temperature of the adlayer (see below) the $\mu'$ quantity could be also estimated at the nonequilibrium temperature. By means of the $\mu'$ expression and neglecting the rate of the reverse process, eqn.1 becomes

$$r = \sum_{i,j,k} w_{i_\alpha j_\lambda, k} p_{i_\alpha} p_{j_\lambda} n_A n_B = \left( \frac{e^{-\beta_s(\overline{E}_A+\overline{E}_B)} \sum_{i,j,k} w_{i_\alpha j_\lambda, k} p_{i_\alpha} p_{j_\lambda}}{e^{\sum_j p_{j_\lambda} \ln p_{j_\lambda}} e^{\sum_i p_{i\alpha} \ln p_{i\alpha}}} \right) e^{\beta_s(\mu'_A + \mu'_B)}, \tag{8}$$

to be compared with eqn.6 (for $A_r \to -\infty$). In the case of thermalized adlayer the temperature of the system is $T_s$ and since $p_{i_\alpha} = \dfrac{e^{-\beta_s E_{i_\alpha}}}{q_s}$ in this case, eqn.8 actually reduces to eqn.6 ($q_s = q(T_s)$) [1].

The temperature of the adlayer can be defined by employing the concept of "generalized heat" [25,26] through the relationship

$$d\overline{E} = Tds \equiv -k_B T \sum_i dp_{i_\alpha} \ln p_{i_\alpha}, \tag{9}$$

where the number of adatoms and the area of the surface are taken as constants and $d\overline{E} = \sum_i E_{i_\alpha} dp_{i_\alpha}$. Eqn.9 is to be considered as the definition of temperature in terms of distribution function (d.f.) and energy spectrum of the single particle. For instance, in the case of an equilibrium adlayer $p_{i_\alpha} = \left( \sum_i e^{-\beta_s E_{i_\alpha}} \right)^{-1} e^{-\beta_s E_{i_\alpha}}$ and eqn.9 gives $T = T_s$. To make the discussion concrete, let us consider the case of a vibrational harmonic ladder where the levels are overpopulated with respect to the Boltzmann distribution computed at the temperature of the surface, $T_s$. To simplify the discussion, from now on the vibrational states are assumed to be non degenerate, i.e. the α index is dropped in the sums.

In order to get a more manageable expression for $T$, the d.f. is rewritten here in terms of the "overpopulation factor", $f_i = \dfrac{\frac{n_i}{n_0}}{\left(\frac{n_i}{n_0}\right)_B} = \dfrac{n_i}{n_0} e^{\beta_s E_i}$, that is the ratio between the relative population of the actual system and the relative population of the Boltzmann distribution computed at the same $n_0$ value. The energy of the ground level is set equal to zero ($E_0 = 0$). Accordingly, $p_i = \dfrac{n_i}{n} = \dfrac{n_0}{n} e^{-\beta_s E_i} f_i$ and eqn.9 gives rise to $\sum_i dp_i E_i = -k_B T \sum_i dp_i [\ln f_i - \beta_s E_i]$, namely

$$\frac{1}{T_s} - \frac{1}{T} = \frac{k_B \sum_i \ln f_i \, dp_i}{\sum_i E_i \, dp_i}, \tag{10a}$$

---

[1] In the present modeling the surface is considered as a thermal reservoir whose temperature is independent of the reactive process.

where $p_0 = \dfrac{n_0}{n}$ and the condition $\sum_i dp_i = 0$ has been used. Clearly, the case $f_i = \text{constant}$ reduces to the Boltzmann distribution function, for $f_0 = 1$ always holds true. Furthermore, $dp_i = e^{-\beta_s E_i} f_i dp_0 + p_0 e^{-\beta_s E_i} df_i$. The conjecture is now made according to which the variation of the overpopulation is a one parameter function of the form $df_i = \dfrac{\partial f_i}{\partial \xi} d\xi$. Under these circumstances and using the constraint above one obtains $g_i = \dfrac{dp_i}{dp_0}$ as

$$g_i = \left[ e^{-\beta_s E_i} f_i - \dfrac{e^{-\beta_s E_i} \frac{\partial f_i}{\partial \xi}}{p_0 \sum_{i=1}^{v^*} e^{-\beta_s E_i} \frac{\partial f_i}{\partial \xi}} \right] \quad \text{for } i \neq 0 \tag{10b}$$

$$g_0 = 1$$

with $v^*$ being the vibrational quantum number of the upper bound level, $p_0^{-1} = \sum_{i=0}^{v^*} e^{-\beta_s E_i} f_i$ and $\sum_{i=0}^{v^*} g_i = 0$. From eqn.10a the following relation is obtained for the vibrational temperature of the adlayer

$$\dfrac{1}{T_s} - \dfrac{1}{T} = k_B \dfrac{\sum_{i=1}^{v^*} g_i \ln f_i}{\sum_{i=1}^{v^*} g_i E_i}. \tag{11}$$

Two cases are worthy of note. In the first the overpopulation factor is independent of quantum number according to the expression $f_i = 1 + (1 - \delta_{i,0})\xi$ where $\delta_{i,0}$ is the Kronecker delta, and $\xi$ is a positive parameter. For a harmonic vibrational ladder $E_v = vE_{01}$, $E_{01}$ being the energy spacing of the ladder, and eqn.11 eventually becomes,

$$\dfrac{1}{T_s} - \dfrac{1}{T} = \dfrac{k_B}{E_{01}} (1-\gamma) \ln(1+\xi) \tag{12}$$

where $\gamma = e^{-\beta_s E_{01}}$ and the case $\beta_s E_{01} > 1$ is considered (see the appendix). It is apparent from eqn.12 that in this case the temperature scales logarithmically on the nonequilibrium parameter, $\xi$.

The second example concerns the d.f. previously derived in refs.[17,27] in the case of prevailing vibrational quantum exchange among adatoms. The d.f. reads,

$$p_v = \frac{n_0}{n} e^{-\beta_s E_v} e^{v\xi}, \qquad (13)$$

where $f_v = e^{v\xi}$ is the overpopulation factor and $\xi$ is a positive constant. For the harmonic vibrational ladder, eqns.11,13 give

$$\frac{1}{T} = \frac{1}{T_s} - \frac{\xi k_B}{E_{01}}. \qquad (14)$$

Alternatively, eqn.14 defines the nonequilibrium parameter in terms of $T$: $\xi = \left(1 - \frac{T_s}{T}\right)\beta_s E_{01}$. The details of the derivation of eqn.14 are reported in Appendix 1. It is worth noticing that, differently to the previous one, in this case the ratio $T_s/T$ scales linearly with $\xi$.

Using the expression of $g_i$ the vibrational temperature, eqn.11, takes the alternative form

$$\frac{T}{T_s} = \frac{\beta_s \sum_i g_i E_i}{\sum_i g_i (\beta_s E_i - \ln f_i)}, \qquad (15)$$

which holds true provided that $f_i < e^{\beta_s E_i}$. In fact, for the d.f. eqn.13 this constraint implies $\xi < \beta_s E_{01}$ in agreement with the above result.

*2-Application to adatom recombination at steady state*
*2-1 Vibrational temperature of the adlayer*

The aim of this section is to compute the vibrational temperature and the entropy of the adatoms in the case of diatom formation, $A_2$, at the steady state. The model takes into account several reaction channels where adatoms in vibrational levels $n$ ($A_n$) and $m$ ($A_m$) react to produce a molecule. As far as the energy transfer is concerned, only energy exchange between adatom and the solid lattice are taken into account. A reaction channel is open for $E_m + E_n \geq E^{\#}$ where $E^{\#}$ is the activation energy and $E_n$ the adatom vibrational energy (fig.1). Owing to the greater occupation number of the ground level, contributions of reaction channels involving this level are considered in the computation, only. Under these circumstances the d.f. is given by [20],

$$f_{\kappa+j} = \frac{(1+\rho)^j}{\gamma^j}\left[\eta + \frac{\rho+(1-\gamma)z^j}{(1+\rho-\gamma)}\right] \quad 1 \leq j \leq v^*-\kappa \tag{16a}$$

$$f_j = 1 + \frac{\gamma^\kappa}{\gamma^j}\eta \quad 0 < j \leq \kappa \tag{16b}$$

where $\rho = w\theta_0/k$, $z = \gamma(1+\rho)^{-1}$, with $w$ being the rate constant for diatom formation, $k$ the rate constant for the loss of a vibrational quantum to the solid, and $\theta_0$ the occupation number (in monolayer units) of the ground level. Furthermore, $\eta = \frac{\rho(1+\rho)^{v^*-\kappa}}{2-(1+\rho)^{v^*-\kappa+1}}$ where $E_\kappa = E^{\#}$ and $E_{v^*}$ is the energy of the upper bound vibrational level. It is worth noticing that $\rho$ is an important quantity of the model for it determines the displacement of the d.f. from the Boltzmann one. The reaction rate, $\Phi$, is given by $\Phi = \Phi_B \frac{\eta}{\rho}$ and in the limit $\rho \to (2^{1/(v^*-\kappa+1)}-1)$ it becomes orders of magnitude higher than the value computed for the Boltzmann distribution at the surface temperature: $\Phi_B = 2w\theta_0^2\gamma^\kappa$. Also, $\lim_{\rho \to 0}\Phi = \Phi_B$. It is worth stressing, however, that the condition $\Phi \cong \Phi_B$ does not necessarily implies that the adlayer has thermalized with the surface. In fact, according to eqn.16 the Boltzmann distribution function is recovered only in the case of a *non-reactive* adlayer, for in this case the overpopulation factors - of *the whole ladder* - are negligible when compared to unity. For high exoergic adsorption and/or low surface temperature this condition is fulfilled, in general, for $w \to 0$, i.e. for $\Phi_B \to 0$.

In the following we limit the analysis to the case in which $\rho$ is lower than unity. Under these circumstances the d.f. eqn.16 takes the approximate simple form

$$f_i = 1 + \frac{\xi}{\gamma^i} \quad \kappa < i \leq v^* \tag{17a}$$

$$f_i = 1 + \frac{\xi}{2\gamma^i} \quad 0 < i \leq \kappa, \tag{17b}$$

where the parameter $\xi$ is proportional to the ratio between reaction rate and rate coefficient for energy disposal to the solid: $\xi = \frac{\Phi}{k\theta_0}$. In fact, the d.f. eqn.17 is typical of reactive adlayer at steady state. In particular, it also holds for reaction channels involving adatoms in the same vibrational level although, in this instance, a multiplicative factor enters the definition of $\xi$ [15]. Accordingly, in what it follows the general form of the vibrational distribution, eqn.17, will be employed. In fig.2 the typical trend of the overpopulation factor given by eqn.17 is reported for $\gamma = 0.05$ and $\xi = 10^{-14}$, $\xi = 10^{-16}$, values which are representative of real systems. In the same figure the overpopulation factor of the distribution derived by Treanor et al in ref. [27] (eqn.13) is also displayed for comparison.

Use of eqn.11 leads to the following expression for the vibrational temperature of the adlayer (see also Appendix 2)

$$\frac{T_s}{T} = 1 - \frac{2}{c\beta_s E_{01}} \left( \sum_{i=1}^{\kappa} \ln f_i + 2 \sum_{i=\kappa+1}^{v^*} \ln f_i \right) \tag{18}$$

with $c(\kappa, v^*) = [2v^*(v^*+1) - \kappa(\kappa+1)]$. Since $\frac{1}{2}D_{A_2} + E_a \cong v^* E_{01}$, with $D_{A_2}$ being the dissociation energy of the molecule, the term $c(\kappa, v^*)$ in eqn.18 can be rewritten as $c(v^*, D) = -2v^{*2} + 4v^*D - D(D-1)$ where $D = \frac{D_{A_2}}{E_{01}}$ and $\frac{1}{2}D < v^* < D$ is assumed. The activation energy of the reaction therefore reads $E^\# = 2E_a = (2v^* - D)E_{01}$ where $E_a$ is the adsorption energy (fig.1).

The contribution of the upper bound level to the vibrational temperature of the adlayer is attained by retaining in eqn.18 only the term at $i = v^*$ according to

$$\left(\frac{T_s}{T}\right)_{v^*} = 1 - \frac{4}{\beta_s E_{01} c} \ln\left(1 + \frac{\xi}{\gamma^{v^*}}\right). \tag{19a}$$

For an ample class of exoergic reactions on catalytic surfaces the constraint $\gamma^{v^*} \ll \xi < 1$ is fulfilled, and eqn.19a becomes

$$\left(\frac{T_s}{T}\right)_{v^*} \cong \left(1 - \frac{4v^*}{c(v^*,D)}\right) + \frac{4}{\beta_s E_{01}} \frac{1}{c(v^*,D)} |\ln \xi|. \tag{19b}$$

Eqn.19b allows us to determine the upper bound of the contribution of level $v^*$ to the vibrational temperature according to $1 \leq \left(\frac{T}{T_s}\right)_{v^*} < \left(1 - \frac{4v^*}{c(v^*,D)}\right)^{-1}$ which is a function of both adsorption energy of the adatom and binding energy of the molecule. Eqns.19a and 19b are important for they make possible to characterize the non-equilibrium state of the system. In fact, the displacement from equilibrium of the vibrational level -namely the overpopulation- is expected to be greater for $i = v^*$. The behaviour of eqn.19a, as a function of the activation energy, is reported in fig.3a for several values of $D_{A_2}$ and at given values of $\xi$ and $\beta_s E_{01}$. In particular, for H recombination on metals $D \approx 36$ (at $E_{01} \cong 0.12 \text{eV}$) and $\xi$ is of the order of magnitude of $\xi \cong 10^{-14} \text{s}^{-1}$. The non linear trend of $(T_s/T)_{v^*}$ on adsorption energy (in fig.3a) is mainly due to the $c(v^*,D)$ function which depends on the energetics of the process. A change in the reaction rate shifts the curve along the $(T_s/T)_{v^*}$ axis, only. This is highlighted in panel b) where the $(T_s/T)_{v^*}$ is plotted as a function of the control parameter $\xi = \frac{\Phi}{k\theta_0}$ for several values of $D$, activation energy and $\beta_s E_{01}$. For H recombination on metals the interval of the experimental values of $\xi = \frac{\Phi}{k\theta_0}$ is also marked in the figure. Moreover, for a given system $(T_s/T)_{v^*}$ increases with reaction rate. In turn, shifting from one system to another, (i.e. changing the $v^*, D$ couple) the temperature exhibits the complex behavior brought about by the non linear term $c(v^*,D)$. With reference to fig.3b, at the lowest binding energy ($D=10$) the adlayer has thermalized (at $\beta E_{01} = 6$) provided $\xi < 10^{-17}$, while at $\beta E_{01} = 8$ its vibrational temperature can be higher than that of a system which is more exoergic, depending on $\xi$. In general, in the case of highly exoergic reaction and/or sufficiently low substrate temperature, for realistic values of $\xi$ the adlayer has not thermalized with the surface. The minimum values $(T_s/T)_{v^*,\min}$, as defined by eqn.19b, are displayed in fig.3c. These values are characteristic of each 'reaction-catalyst' system, for they depend on the energetics, only. Values of $(T_s/T)_{v*,\min}$ in fig.3c span the ample range 0.6-0.9; as a general trend the higher $D$ the higher this figure.

To include in the computation of $T_s/T$ the contribution of all levels, the sum over the logarithms of the overpopulation factors has to be carried out. This can be accomplished, analytically, by retaining in the sum only the leading terms, for which $\frac{\xi}{\gamma^i} \gg 1$. This condition is usually well satisfied for energies greater than the activation energy, i.e. for $E_i > E_\kappa$. According to the derivation reported in Appendix 2, eqn.18 becomes

$$\frac{T_s}{T} = \frac{\kappa(\kappa+1)}{c} + \frac{4(v*-\kappa)}{c\beta_s E_{01}}|\ln\xi| \quad (20)$$

to be compared with eqn.19b. Computations of the vibrational temperature according to eqn.20 are reported in fig.4 as a function of $\xi$ and for several values of $D$, and $v*$. This figure shows that the contribution of the level with $i > \kappa$ is important in determining the vibrational temperature; the contribution of the whole ladder is, on average, two times that of the upper bound level. On one hand, for a given system and at a given $\xi$ the vibrational temperature of the adlayer decreases with the substrate temperature, that is the adlayer is more "hot" the lower $T_s$. On the other hand, the vibrational temperature increases with $D$, namely the system is more "hot" the higher the energy released by the reaction.

Above we have defined the quantity $\Phi_B = 2w\theta_0^2\gamma^\kappa$ as the "hypothetical" recombination rate for a Boltzmann d.f., at the temperature of the surface, and at the actual value of $\rho = \frac{w\theta_0}{k}$. The behavior of $(T_s/T)_{v*}$ for $\Phi = \Phi_B$ is displayed in fig.5a as a function of $E^\#$ for two values of $D$. In the same figure it is also reported the $(T_s/T)_{v*}$ ratio for parameter values typical of H recombination on metals ($D \approx 36$ and $\xi \approx 10^{-14}$) where adsorption energies are in the interval 0.3-0.6eV and $E^\#/E_{01}$ ranges between 5 and 12 [21]. One notices that for $\xi = 2\times 10^{-14}$ the recombination rate are several orders of magnitude higher than $\Phi_B$, although this entails a vibrational temperature increase of a few per cent only (fig.4a). This is due to the strong non-linear dependence of reaction rate on adatom vibrational temperature. This aspect is further discussed in sect.2.3.

The entropy of the adlayer is computed through eqn.7 by means of the d.f. eqn.17. Performing the sums of the arithmetic series and retaining the leading term of the logarithmic contribution, one ends up with (see also Appendix 3)

$$s = s_0 - k_B \ln n + k_B p_0 \xi \beta_s E_{01} \frac{c(v^*,D)}{4} \frac{T_s}{T}, \qquad (21a)$$

where $s_0 = -k_B \ln p_0 + k_B p_0 \beta_s E_{01} \frac{\gamma}{(1-\gamma)^2}$. For $\xi = 0$ (i.e. $\Phi=0$), $T = T_s$, $p_0 = q_s^{-1}$ and the usual expression of the entropy is obtained. Besides, for $\beta_s E_{01}$ values much higher than unity $p_0 = \left(\sum_n \gamma^n f_n\right)^{-1} \cong 1-\gamma$. The excess entropy (with respect to that of the thermalized adlayer) reads

$$s^{ex} = s_0^{ex} - k_B \ln(n/n_B), \qquad (21b)$$

where $s_0^{ex} = k_B p_0 \xi \beta_s E_{01} \frac{c(v^*,D)}{4} \frac{T_s}{T}$ and $n_B$ the adsorbed quantity at equilibrium. The ratio between the excess entropy of the adlayer, $s_0^{ex}$, and that of the same system at $\Phi = \Phi_B$ is displayed in fig.5b as a function of $\xi$. Vibrational temperatures of the adlayer are also reported. The plot indicates that the excess entropy scales almost linearly on $\xi$, i.e. on recombination rate.

*2-2 Equilibrium condition and linear response equation*

This section is devoted to the equilibrium condition of the adlayer that is actually achieved, as discussed in the previous section, at $\Phi = 0$ ($T = T_s$). So far we have dealt with the adlayer, only. In order to study the system at equilibrium, however, also the species in the gas phase have to be considered. In fact, at the steady state the reaction rate (expressed as the number of adatoms that recombine per unit of time) is linked to the process of adsorption-desorption through the expression

$$2r \equiv \Phi = F(1-\theta) - v_0 p_{v^*} \theta, \qquad (22)$$

where $F = \dfrac{P}{N\sqrt{2\pi m k_B T_s}}$ is the flux of incoming gas atoms at the surface, $P$ the gas pressure, $N$ the surface density of adsorption sites, $m$ the mass of the adatom, $\nu_0$ the rate constant for desorption and $\theta$ is the total surface coverage. By exploiting the definition of gas and adatom chemical potential, $\mu_g = \mu_g^0 + k_B T_s \ln P/P^0$ and $\mu' = \mu'^0 + k_B T_s \ln \theta$ respectively, eqn.22 can be rewritten as $\Phi = \alpha(1-\theta)e^{\beta_s(\mu_g - \mu_g^0)} - \nu_0 p_{v^*} e^{\beta_s(\mu' - \mu'^0)}$ with $\alpha = \dfrac{P_0}{N\sqrt{2\pi m k_B T_s}}$. Alternatively, exploiting the equilibrium condition, $\Phi = 0$, $\theta = \theta_{eq}$, eqn.22 becomes

$$\Phi = \dfrac{(1-\theta)}{(1-\theta_{eq})} \nu_0 p_{v^*,eq} e^{\beta_s(\mu'_{eq} - \mu'^0_{eq})} - \nu_0 p_{v^*} e^{\beta_s(\mu' - \mu'^0)}$$

$$= \dfrac{(1-\theta)}{(1-\theta_{eq})} \nu_0 p_{v^*,eq} e^{\beta_s(\mu'_{eq} - \mu'^0_{eq})} \left[ 1 - \dfrac{(1-\theta_{eq})}{(1-\theta)} \dfrac{p_0}{p_{0,eq}} f_{v^*} e^{\beta_s(\mu' - \mu'_{eq} - (\mu'^0 - \mu'^0_{eq}))} \right].$$

(23)

In the limiting case of very small departure from equilibrium $\theta \cong \theta_{eq}$ and the thermodynamic equation of motion is formally obtained as

$$\Phi \approx -\dfrac{\nu_0 \gamma^{v^*} \theta_{eq}}{k_B T_s}(\mu' - \mu'_{eq}),$$

(24)

with the linear response coefficient $L = \dfrac{\nu_0 \theta_{eq}}{k_B T_s} e^{-\beta_s(\frac{1}{2}D_{A_2} + E_a)}$. In the eqn.23 the difference $\mu'^0 - \mu'^0_{eq} = (\bar{E} - \bar{E}_{eq} - T_s s_0^{ex})$ has been neglected in the exponential function. In fact, in the expression $\bar{E} - \bar{E}_{eq} \approx (f_1 - 1)\gamma E_{01}$, $\gamma < 1$ and $f_1 - 1 \ll 1$, while the excess entropy is of the order of magnitude of $\xi$ that is in the range $10^{-15} - 10^{-12}$. However, the validity of the linear approximation is linked to the value of the coverage at steady state and, in turn, to the energetics of the reaction as discussed in the next section. Besides, as maintained in [24], catalytic processes usually take place under condition far from equilibrium where the linear response theory does not apply.

*2-3 Surface coverage and reaction rates*

On the basis of the present approach, under nonequilibrium conditions the fractional surface coverage is a function of recombination rate. Also, the relationship between these two quantities is, in general, non-linear. Specifically, both $\Phi$ (i.e. $\xi$) and $\theta$ can be determined by solving the system of equations eqn.22 and eqn.8, where the equilibrium condition implies $w = 0$. For the d.f. eqn.17 with activation energy $E^{\#} = E_{\kappa}$, the system can be solved analytically by employing reasonable approximations. Using eqn.17 in eqn.8 the reaction rate is computed as

$$\Phi = w' p_0^2 \theta^2 \left[ \left( \gamma^{\kappa} + \frac{\xi}{2} \right) + \sum_{j=1}^{v^*-\kappa} \left( \gamma^{j+\kappa} + \xi \right) \right] \cong w' p_0^2 \theta^2 \left[ \gamma^{\kappa} + c'\xi \right] \quad \text{where} \quad c' = \frac{1}{2}[2(D - v^*) + 1],$$

$w' = 2w$ and $p_0 = \left( \sum_n \gamma^n f_n \right)^{-1} \cong 1 - \gamma$ ($\gamma \gg \xi$). Consequently, since by definition $\Phi = k\xi\theta$ one gets

$$\theta \cong \frac{k}{w'} \left( \frac{\xi}{e^{-\beta_s E^{\#}} + c'\xi} \right), \tag{25}$$

where $p_0 \cong 1$ was employed. By means of eqns.22,25 the following equation is obtained for $\xi$, eventually

$$\bar{k}\xi'^2 + (\lambda - \tfrac{w'}{k} c')\xi' - \tfrac{w'}{k} = 0, \tag{26a}$$

where $\xi' = \xi e^{\beta_s E^{\#}}$, $\bar{k} = \frac{1}{F}[k + v_0]e^{-\beta_s E^{\#}}$, $\lambda = \frac{1}{\theta_{eq}} = 1 + \frac{v_0 e^{-\beta_s E_{v^*}}}{F}$, and $p_0 \cong 1$ was again assumed. The solution for $\xi'$ reads,

$$\xi' = \frac{1}{2\bar{k}} \left( -(\lambda - \tfrac{w'}{k} c') + |\lambda - \tfrac{w'}{k} c'| \sqrt{1 + \frac{4 \tfrac{w'}{k} \bar{k}}{(\lambda - \tfrac{w'}{k} c')^2}} \right). \tag{26b}$$

One notices that in the limit $\tfrac{w'}{k} \to 0$, $\xi' \approx \frac{w'}{k}\frac{1}{\lambda}$ and eqn.25 gives $\theta \cong \tfrac{1}{\lambda} = \theta_{eq}$. For $(\lambda - \tfrac{w'}{k} c') \to 0$ $\xi' \approx \left( \frac{w'}{k\bar{k}} \right)^{1/2}$ and the surface coverage (eqn.25) becomes $\frac{1}{\theta} = \frac{1}{\theta_{eq}} + \left( \frac{\bar{k}}{\theta_{eq} c'} \right)^{1/2}$.

On the other hand, in the case of high desorption energy of the adatoms ($\beta_s E_{v^*} \gg 1$), high reaction rates are attained for $\frac{w'}{k}c' > \lambda \cong 1$ and $\bar{k} < 1$. In fact, under these circumstances $\xi' \approx \frac{1}{\bar{k}}\left(\frac{w'c'}{k} - 1\right)$ where $c' > 1$ and $\lambda \cong 1$ is used. In addition $\frac{1}{\theta} = \frac{w'}{k}\left[\frac{\bar{k}}{\left(\frac{c'w'}{k} - 1\right)} + c'\right] \approx \frac{w'c'}{k}$ that is greater than $\frac{1}{\theta_{eq}} = \lambda \cong 1$. Consequently, the ratio $\frac{\Phi}{\Phi_B}$, being $\Phi_B = w'\theta^2 e^{-\beta_s E^\#}$ the reaction rate for the Boltzmann distribution, reads

$$\frac{\Phi}{\Phi_B} = \frac{c'}{\bar{k}}\left(\frac{c'w'}{k} - 1\right) \approx \frac{F}{(k+v_0)}e^{\beta_s E^\#} \tag{27}$$

that can be several orders of magnitude higher than unity. Interestingly, the value of this ratio is dictated by the relative magnitude of the rate constant for energy disposal (desorption) and the activation energy containing term. In the case of highly exoergic reactions the conditions that validate eqn.27 are usually satisfied. For $k \approx 10^{13} \text{s}^{-1}$ the constraint $\bar{k} < 1$ is verified for $\beta_s E^\# > 29$. In particular, in the case of H recombination on metals $F \approx 0.1 \text{s}^{-1}$, $k$ and $v_0$ are of the order of $10^{13} \text{s}^{-1}$ and for $E_{01} \approx 0.12 \text{eV}$ and $E_a \approx 0.3 \text{eV}$ one gets $c' = \frac{1}{2}\left[\frac{D_{H_2} - 2E_a}{E_{01}} + 1\right] \approx 16$ [21]. At $T = 150\text{K}$ one obtains $\bar{k} \approx 10^{-6}$ and $\frac{1}{\theta} \cong \frac{c'w'}{k}$ is therefore a very good approximation.

It is instructive to estimate the vibrational temperature of the adlayer for the kinetic model discussed so far. Use of eqn.20 gives

$$\frac{T_s}{T} = \frac{\kappa(\kappa+1)}{c} + \frac{4(v^* - \kappa)}{c}\left[\kappa - \frac{1}{\beta_s E_{01}}\left(\ln\frac{w'}{k} + \ln\theta + \ln\frac{\Phi}{\Phi_B}\right)\right], \tag{28}$$

where the logarithmic dependence of $T_s/T$ on reaction rate and surface coverage has been made explicit. In terms of rate coefficients the temperature becomes, $\frac{T_s}{T} = \frac{\kappa(\kappa+1)}{c} - \frac{4(v^* - \kappa)}{c\beta_s E_{01}}\ln\left[\left(c'\frac{w'}{k} - 1\right)\frac{F}{k+v_0}\right]$ where the argument of the logarithmic term is expected to be lower than one. For instance, for hydrogen recombination at $\theta \approx 0.25$, using the quantities above one gets $\frac{T_s}{T} \cong 0.3$ and $\Phi/\Phi_B \cong 10^7$; in other words, a factor of three variation of the temperature of the adlayer entails a reaction rate enhancement of several orders of magnitude.

**Conclusions**

A thermodynamic approach has been developed for determining the vibrational temperature of adatoms during diatom formation at the steady state. The model exploits the definition of generalized heat, together with the definition of nonequilibrium entropy, which requires the knowledge of the energy distribution function of the ad-atoms. These d.f. have been computed by means of kinetic rate equations. It is shown that, provided the reaction rate is different from zero, the temperature of the adlayer is higher than the temperature of the substrate to an extent that depends, logarithmically, on both coverage and reaction rate normalized to the reaction rate of a fully thermalized adlayer. Typical figures for H recombination on metals indicates that a vibrational temperature increase of a factor of three brings about a reaction rate enhancement of several orders of magnitude. On the other hand, the ad-layer temperature exhibits a complex non-linear dependence on adsorption energy and binding energy of the molecule.

The present approach is discussed in connection with the linear equations of the nonequilibrium thermodynamics and the response coefficient determined. In particular, it has been shown to be proportional to the rate constant for adatom desorption.

The definition of the adlayer temperature here proposed is shown to be consistent with the results previously attained in the literature in the case of prevailing vibrational quantum exchange among molecules in the gas phases.

**Appendix 1**

In this appendix the computations of eqn.11 for the two nonequilibrium distribution functions discussed in the text, are reported. The evaluation of eqn.11 requires the determination of the sum $\sum_{i=1}^{v^*} E_i g_i$, that is

$$\sum_{i=1}^{v^*}(iE_{01})g_i = E_{01}\left[(1+\xi)\sum_{i=1}^{v^*}ie^{-\beta_s E_{01}i} - \frac{\sum_{i=1}^{v^*}ie^{-\beta_s E_{01}i}}{p_0\sum_{i=1}^{v^*}e^{-\beta_s E_{01}i}}\right] \quad (A1)$$

$$\cong E_{01}\left[(1+\xi)\frac{\gamma}{(1-\gamma)^2} - \frac{1}{p_0(1-\gamma)}\right]$$

with $\gamma = e^{-\beta_s E_{01}}$. Since, $\sum_{i=0}^{v^*} p_i = 1$ one gets $\frac{1}{p_0} = 1 + (1+\xi)\frac{\gamma}{1-\gamma}$ which can be used in eqn. A1 to eliminate $p_0$. Moreover, $\sum_{i=1}^{v^*} g_i \ln f_i = \sum_{i=0}^{v^*} g_i \ln(1+\xi) - g_0 \ln(1+\xi) = -\ln(1+\xi)$, where the condition $\sum_i g_i = 0$ has been exploited. Accordingly,

$$\frac{1}{T_s} - \frac{1}{T} = k_B \frac{\ln(1+\xi)}{E_{01}(1-\gamma)^{-1}}, \quad (A2)$$

that is eqn.12.

For the distribution function, $f_\nu = e^{\nu\xi}$, it is obtained

$$\sum_{i=1}^{v^*} g_i E_i = \left(\sum_{i=1}^{v^*} e^{-\beta_s E_i} f_i E_i - \frac{\sum_{i=1}^{v^*} e^{-\beta_s E_i} E_i \frac{\partial f_i}{\partial \xi}}{p_0 \sum_{i=1}^{v^*} e^{-\beta_s E_i} \frac{\partial f_i}{\partial \xi}}\right) = E_{01}\left(\sum_{i=1}^{v^*} ie^{-(\beta_s E_{01}-\xi)i} - \frac{\sum_{i=1}^{v^*} i^2 e^{-(\beta_s E_{01}-\xi)i}}{p_0 \sum_{i=1}^{v^*} ie^{-(\beta_s E_{01}-\xi)i}}\right),$$

$$\sum_{i=1}^{v^*} g_i \ln f_i = \left(\sum_{i=1}^{v^*} e^{-\beta_s E_i} f_i \ln f_i - \frac{\sum_{i=1}^{v^*} e^{-\beta_s E_i} \frac{\partial f_i}{\partial \xi} \ln f_i}{p_0 \sum_{i=1}^{v^*} e^{-\beta_s E_i} \frac{\partial f_i}{\partial \xi}}\right) = \xi\left(\sum_{i=1}^{v^*} ie^{-(\beta_s E_{01}-\xi)i} - \frac{\sum_{i=1}^{v^*} i^2 e^{-(\beta_s E_{01}-\xi)i}}{p_0 \sum_{i=1}^{v^*} ie^{-(\beta_s E_{01}-\xi)i}}\right),$$

that is

$$\sum_i g_i \ln f_i = \frac{\xi}{E_{01}} \sum_i g_i E_i, \quad (A3)$$

from which eqn.14 is derived.

**Appendix 2**

For the d.f. eqn.17 the $g_i$ terms are given by ( $0 < i \leq \kappa$ )

$$g_i = \gamma^i + \frac{\xi}{2} - \frac{\frac{1}{2}}{p_0 \left( \sum_{n=1}^{\kappa} \frac{1}{2} + \sum_{j=1}^{v^*-\kappa} 1 \right)} = \gamma^i + \frac{\xi}{2} - \frac{1}{p_0(\kappa + 2(v^*-\kappa))} = \gamma^i f_i - \frac{1}{p_0 D} \tag{A4}$$

where $D = (2v^* - \kappa)$.

A similar computation shows that for $\kappa < i \leq v^*$

$$g_i = \gamma^i + \xi - \frac{1}{p_0 \left( \sum_{n=1}^{\kappa} \frac{1}{2} + \sum_{j=1}^{v^*-\kappa} 1 \right)} = \gamma^i f_i - \frac{2}{p_0 D} . \tag{A5}$$

It follows:

$$\sum_i g_i \ln f_i = \sum_{i=1}^{\kappa} \left( \gamma^i f_i - \frac{1}{p_0 D} \right) \ln f_i + \sum_{i=\kappa+1}^{v^*} \left( \gamma^i f_i - \frac{2}{p_0 D} \right) \ln f_i$$
$$= \sum_{i=1}^{v^*} \gamma^i f_i \ln f_i - \frac{1}{p_0 D} \left( \sum_{i=1}^{\kappa} \ln f_i + 2 \sum_{i=\kappa+1}^{v^*} \ln f_i \right) \tag{A6}$$

and

$$\sum_i g_i E_i = -\beta_s^{-1} \sum_i g_i \ln \gamma^i = -\beta_s^{-1} \sum_{i=1}^{v^*} \gamma^i f_i \ln \gamma^i - \frac{1}{p_0 D} E_{01} \left( \sum_{n=1}^{\kappa} n + 2 \sum_{n=\kappa+1}^{v^*} n \right)$$
$$= -\beta_s^{-1} \sum_{i=1}^{v^*} \gamma^i f_i \ln \gamma^i - \frac{1}{p_0 D} E_{01} \left[ v^*(v^*+1) - \tfrac{1}{2}\kappa(\kappa+1) \right] \tag{A7}$$
$$= -\beta_s^{-1} \sum_{i=1}^{v^*} \gamma^i f_i \ln \gamma^i - \frac{c}{2 p_0 D} E_{01}$$

Using eqn.11 one eventually gets

$$\frac{T_s}{T}=1+\frac{\sum_{i=1}^{v^*}\gamma^i f_i \ln f_i -\frac{1}{p_0 D}\left(\sum_{i=1}^{\kappa}\ln f_i + 2\sum_{i=\kappa+1}^{v^*}\ln f_i\right)}{\sum_i \gamma^i f_i \ln \gamma^i + \frac{c\beta_s E_{01}}{2p_0 D}} \cong 1-\frac{2}{c\beta_s E_{01}}\left(\sum_{i=1}^{\kappa}\ln f_i + 2\sum_{i=\kappa+1}^{v^*}\ln f_i\right) \quad (A8)$$

where terms of the order of $\gamma^i \ln f_i$ can be neglected compared with the terms in the bracket. In addition, the leading terms in the bracket are those involving vibrational levels with $i>\kappa$ where, in turn, the approximation $\ln f_i \approx \ln\frac{\xi}{\gamma^i}$ is expected to hold for realistic values of $\xi$. Accordingly eqn.A8 becomes

$$\frac{T_s}{T}\cong 1-\frac{4}{c\beta_s E_{01}}\sum_{i=\kappa+1}^{v^*}\left(-\ln\gamma^i + \ln\xi\right)=1-\frac{c-\kappa(\kappa+1)}{c}-\frac{4(v^*-\kappa)}{c\beta_s E_{01}}\ln\xi$$
$$=\frac{\kappa(\kappa+1)}{c}+\frac{4(v^*-\kappa)}{c\beta_s E_{01}}|\ln\xi| \quad (A9)$$

where $\xi \ll 1$.

## Appendix 3

In this appendix we compute the term $s'=-k_B\sum_i p_i \ln p_i$ which enters the expression of the entropy of the adlayer, eqn.7. By using the $p_i$ expression in terms of $f_i$ one gets

$$-\frac{s'}{k_B}=\sum_i p_0 f_i \gamma^i[\ln(p_0\gamma^i)+\ln f_i]=\sum_{i=1}^{\kappa}p_0\left(\gamma^i+\frac{\xi}{2}\right)\ln(p_0\gamma^i)+\sum_{i=\kappa+1}^{v^*}p_0(\gamma^i+\xi)\ln(p_0\gamma^i)+$$
$$+\sum_{i=1}^{\kappa}p_0\left(\gamma^i+\frac{\xi}{2}\right)\ln f_i + \sum_{i=\kappa+1}^{v^*}p_0(\gamma^i+\xi)\ln f_i \quad (A10)$$

that is

$$-\frac{s'}{k_B}=-\frac{s_0}{k_B}+\sum_{i=1}^{\kappa}p_0\frac{\xi}{2}\ln(p_0\gamma^i)+\sum_{i=\kappa+1}^{v^*}p_0\xi)\ln(p_0\gamma^i)+$$
$$+\sum_{i=1}^{\kappa}p_0\left(\gamma^i+\frac{\xi}{2}\right)\ln f_i + \sum_{i=\kappa+1}^{v^*}p_0(\gamma^i+\xi)\ln f_i \quad . \quad (A11)$$

The first two sums give the term $-p_0\xi\beta_s\frac{E_{01}}{4}c$. Moreover, by using eqn.A8 the contribution $p_0\xi\left(\sum_{i=1}^{\kappa}\frac{1}{2}\ln f_i + \sum_{i=\kappa+1}^{v^*}\ln f_i\right)$ can be rewritten in terms of the $T_s/T$ as follows

$$\frac{p_0 \xi}{2}\left(\sum_{i=1}^{\kappa} \ln f_i + 2\sum_{i=\kappa+1}^{\nu^*} \ln f_i\right) = \frac{p_0 \xi}{2}\frac{\beta_s E_{01} c}{2}\left(1 - \frac{T_s}{T}\right). \tag{A12}$$

Therefore, specifying these expressions in eqn.A11 and neglecting the terms of the order of $\gamma^i \ln f_i$ (see also Appendix 2) the entropy is eventually computed as

$$s' \cong s_0 + \frac{1}{4} k_B p_0 \xi \beta_s E_{01} c \frac{T_s}{T}. \tag{A }$$

**Caption of the figures**

Fig.1 Schematic representation of the harmonic vibrational ladder of the adatoms in the 1D adsorption potential well. In the drawing, $\Phi$ is the recombination rate, $\Phi_n$ is the flux of adatoms which leave the n-th vibrational level as diatoms, $E_a$ is the adsorption energy, $D_{A_2}$ is the diatom binding energy, $E_{01}$ the energy spacing of the ladder and $E_\kappa = \kappa E_{01} = E^\#$ the activation energy for recombination.

Fig.2- Typical behavior of the overpopulation factors for the vibrational distribution function employed in sections 2.1 and 2.2. Solid symbols refer to the d.f. eqn.17 at $\gamma = 0.05$, $\kappa = 9$ and $\xi = 10^{-14}$ (squares), $\xi = 10^{-16}$ (diamonds). Open symbols are the distribution function derived in ref.[22], eqn.13, for $\xi = 0.5$.

Fig.3-In panel a) the contribution of the upper bound level to the vibrational temperature is reported as a function of the activation energy $E^\#$, for several values of the diatom binding energy ($D = D_{A_2}/E_{01}$) in the range 10-36. The computations refer to $\Phi/k\theta_0 = 2 \times 10^{-14}$ and $\beta_s E_{01} = 6$. In the figure the behavior of the minimum value of $(T_s/T)_{v*}$ is also displayed for D=26 (dashed line).

Panel b) shows the $(T_s/T)_{v*}$ ratio as a function of the control parameter $\xi = \Phi/k\theta_0$ for several values of D, at $\beta_s E_{01} = 6$ (dashed lines) and $\beta_s E_{01} = 8$ (solid lines). In particular, D=36, v*=21 (open circles); D=26, v*=16 (open squares); D=20, v*=13 (solid triangles) and D=10, v*=6 (solid squares).

Panel c): Minimum values of $(T_s/T)_{v*}$ as a function of activation energy for several values of D in the range 10-36. Since $E^\# = (2v*-D)E_{01}$ the activation energy is in the range $0 < \dfrac{E^\#}{E_{01}} < D$.

Fig.4 Vibrational temperature of the adlayer as a function of $\xi = \Phi/k\theta_0$ computed through eqn.20. Dashed and solid lines refer to $\beta_s E_{01} = 6$ and $\beta_s E_{01} = 8$, respectively. Parameter values are: D=36, v*=21 (open circles); D=26, v*=16 (open squares) and D=20, v*=13 (solid triangles).

Fig.5 Panel a). The temperature ratio $(T_s/T)_{v^*}$ is displayed as a function of activation energy for the "Boltzmann" recombination rate, $\Phi_B$, at $w\theta_0/k = 0.1$, for $D=36$ (solid circles, full line) and $D=20$ (dashed line) at $\beta_s E_{01} = 8$. The computation for $\xi = \Phi/k\theta_0 = 2\times 10^{-14}$ and $D=36$ is also shown (solid diamonds). The arrow indicates the typical value of the activation energy for H recombination on metals. Specifically, $E_a \approx 0.6\text{eV}$, $E_{01} \cong 0.12\text{eV}$ and $D_{H_2} \cong 4.5\text{eV}$.

Panel b). The normalized excess entropy of the adlayer, $s_0^{ex}/s_{0,B}^{ex}$ is shown as a function of recombination rate for $D=36$, $v^*=21$ at $\beta_s E_{01} = 6$ (solid diamonds, full line) and $\beta_s E_{01} = 8$ (solid circles, full line). $s_{0,B}^{ex}$ is the excess entropy at $\Phi = \Phi_B$ ($w\theta_0/k = 0.1$). The computation is performed for $\xi = \Phi/k\theta_0 = 2\times 10^{-14}$ and $n = n_B$. The behaviour of the $T_s/T$ ratio is also displayed as dashed lines for $D=36$, $\beta_s E_{01} = 6$ (solid diamonds) and $\beta_s E_{01} = 8$ (solid circles).

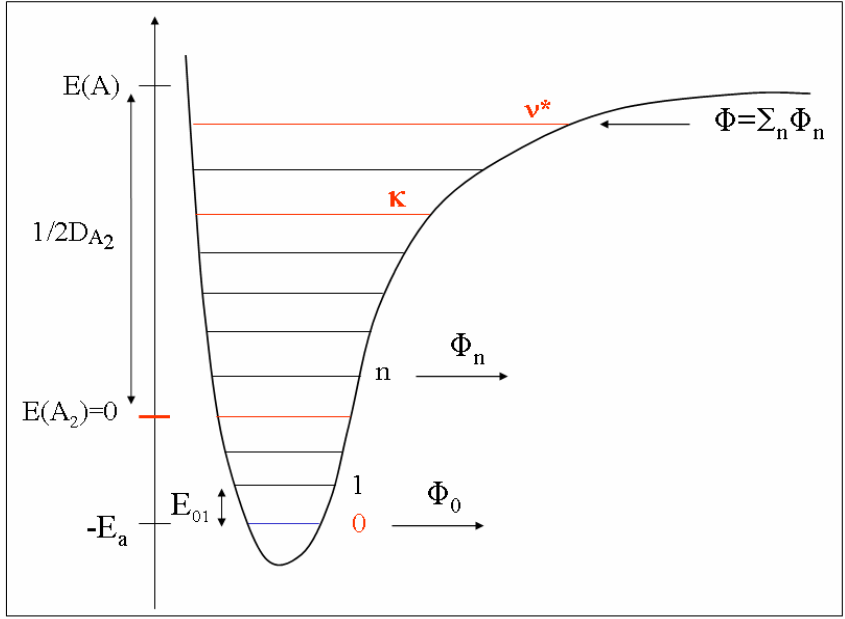

Fig.1

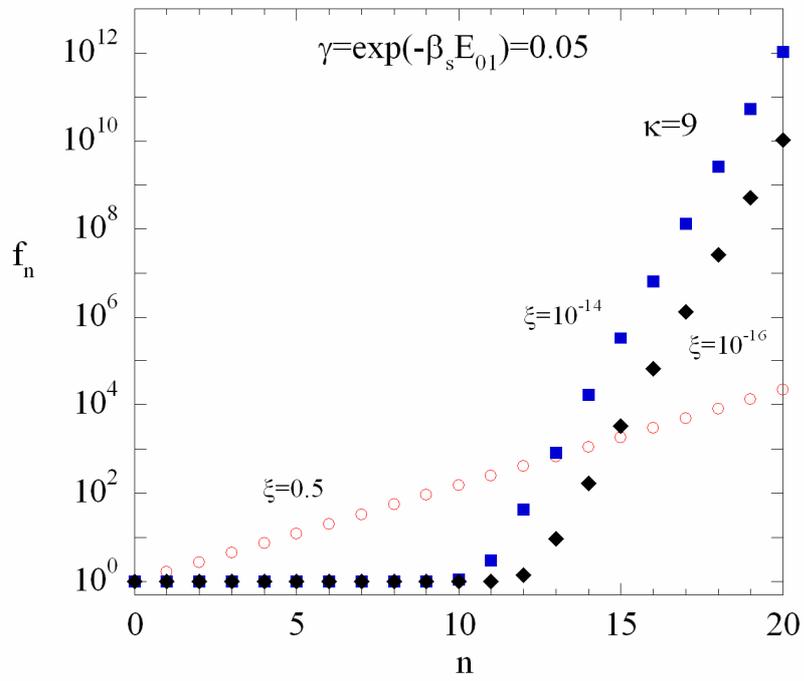

Fig.2

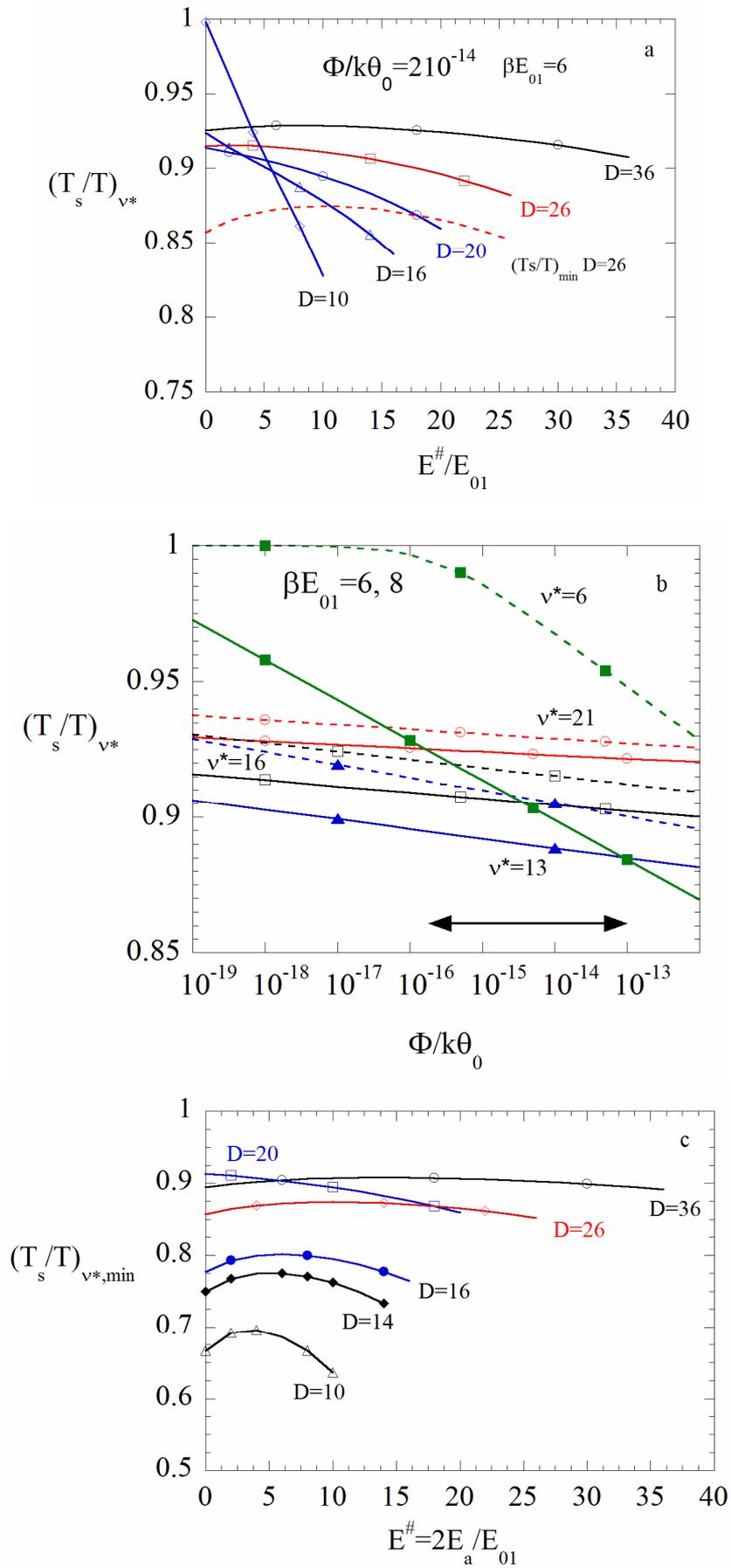

Fig.3

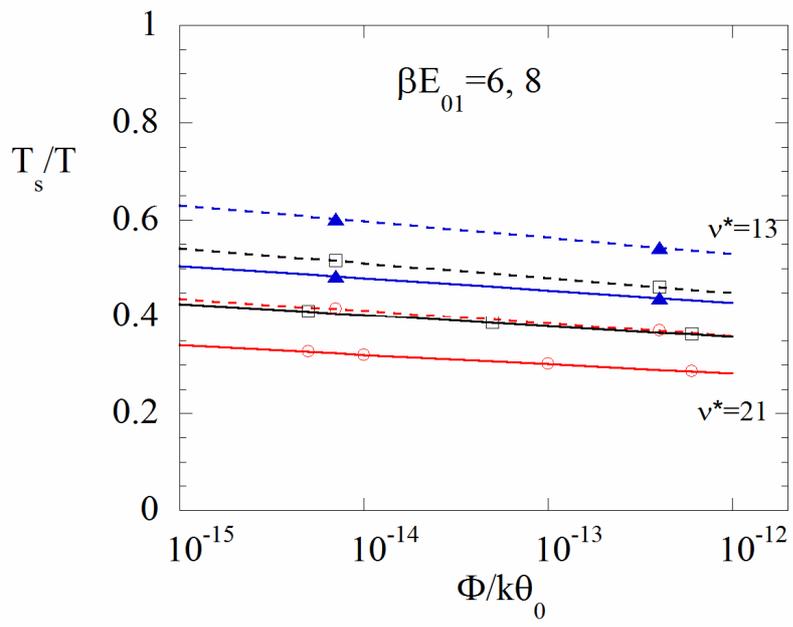

Fig.4

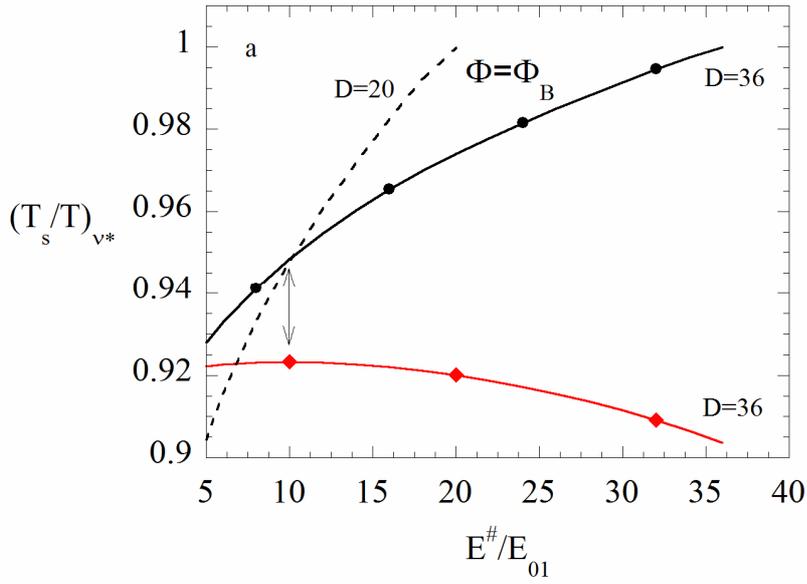

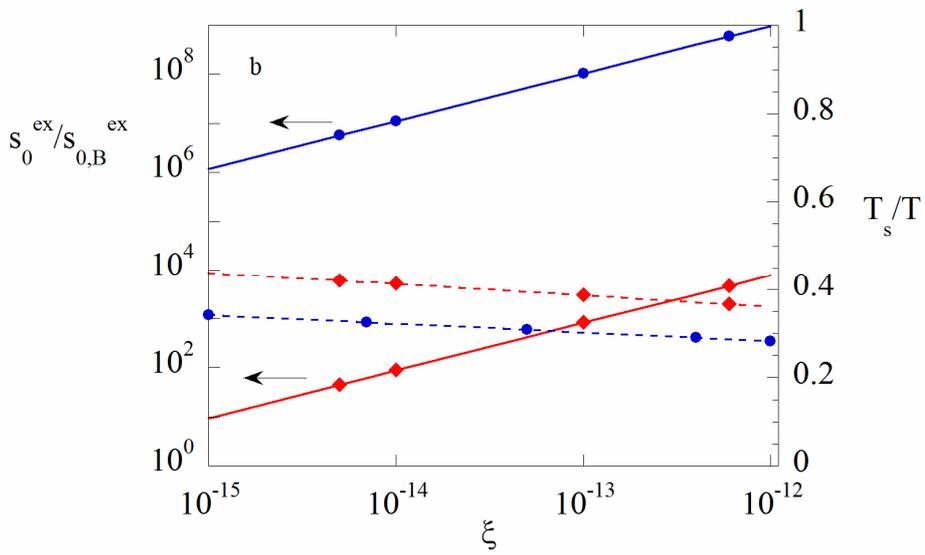

Fig.5